\documentclass[twocolumn,showpacs,preprintnumbers]{revtex4}

\input{epsf}

\def\PL{{\it Phys. Lett.} }
\def\PR{{\it Phys. Rev.} }
\def\PRL{{\it Phys. Rev. Lett.} }

 \def\be{\beta}  
   
\def\th{\theta}

\def\La{\Lambda}

 \def\frac#1#2{{\textstyle{{#1}\over
{#2}}}} 
\def\lsim{\mathrel{\rlap{\lower4pt\hbox{\hskip1pt$\sim$}}
\raise1pt\hbox{$<$}}}
\def\gsim{\mathrel{\rlap{\lower4pt\hbox{\hskip1pt$\sim$}}
\raise1pt\hbox{$>$}}} \def\sqr#1#2{{\vcenter{\vbox{\hrule height.#2pt
\hbox{\vrule width.#2pt height#1pt \kern#1pt \vrule width.#2pt} \hrule
height.#2pt}}}}

\def\beq{\begin{equation}} \def\eeq{\end{equation}}
\def\beqa{\begin{eqnarray}} \def\eeqa{\end{eqnarray}}

\long\def\symbolfootnote[#1]#2{\begingroup
\def\thefootnote{\fnsymbol{footnote}}\footnote[#1]{#2}\endgroup}

\begin{document}

\title{Astrophysical constraints on unparticle-inspired models of gravity}

\author{O. Bertolami\symbolfootnote[4]{Also at {\it Instituto de Plasmas e Fus\~ao Nuclear}, Avenida Rovisco Pais 
1, 1049-001 Lisboa, Portugal.}} \email{orfeu@cosmos.ist.utl.pt}

\author{J. P\'aramos\footnotemark[4]} \email{paramos@ist.edu}

\author{P. Santos\footnotemark[4]} \email{paulo_santos@nfist.ist.utl.pt}

\affiliation{Instituto Superior T\'ecnico, Departamento de
F\'{\i}sica, \\Avenida Rovisco Pais 1, 1049-001 Lisboa, Portugal
}

\date{\today}

\begin{abstract} We use stellar dynamics arguments to constrain the relevant parameters of ungravity 
inspired models. We show that resulting bounds do constrain the parameters of the theory of unparticles, 
as far as its energy scale satisfies the condition $\Lambda_U \geq 1~ TeV$ and $d_U$ is close to unity.

\vskip 0.5cm

\end{abstract}

\pacs{04.20.Fy, 04.80.Cc, 04.25.Nx \hspace{2cm}Preprint DF/IST-5.2009}

\maketitle


\section{Introduction}

It has been remarked that the Standard Model (SM) is likely to be incomplete due to the apparent lack of scale invariant objects, unparticles \cite{georgi}, besides its well-known shortcomings. Implementing 
scale invariance requires considering an additional set of fields with a nontrivial IR fixed point, the Banks-
Zacks (BZ) fields. The interaction between SM and BZ fields occurs through the exchange of particles with a 
large mass scale, $M_*$, written as
\beq {\cal L}_{BZ} = {1\over M_*^k}O_{SM}O_{BZ}~~,
\label{BZ}
\eeq \noindent where $O_{SM}$ is an operator with mass dimension $d_{SM}$ built out of SM fields and 
$O_{BZ}$ is an operator with mass dimension $d_{BZ}$ built out of BZ fields.

At an energy scale $\La_U$ the BZ operators match onto unparticles operators ($O_{U}$) 
and Eq. (\ref{BZ}) matches onto
\beq \label{LagU}
{\cal L}_U={C_U \Lambda_U^{d_{BZ}-d_U}\over M_*^k}O_{SM}O_{U}~~,
\eeq \noindent where $d_U$ is the scaling dimension of $O_U$, which can be fractional, and $C_U$ is a coefficient function.

Considering tensor-type unparticle interactions with the stress-energy tensor of SM states leads to a modification to the 
Newtonian potential $\Phi(r)$, usually referred to as ungravity --- a gravitational potential with a power-law addition \cite{ungravity},
\beq V(r)=-{G_U M \over r}\left[1+\left({R_G \over r}\right)^{2d_u-2}\right]~~,
\label{UPO}
\eeq \noindent where $R_G$ is the characteristic length scale of ungravity,
\beqa 
\label{R_G} R_G & = &{1\over \pi \Lambda_U} \left({M_{Pl} \over M_*}\right)^{1/(d_U-1)} \times \\ \nonumber && \left[ {2(2-
\alpha) \over \pi} {\Gamma(d_U+{1 \over 2})\Gamma(d_U-{1 \over 2}) \over \Gamma(2d_U)}\right]^{1/
(2d_U-2)}~~,
\eeqa 
\noindent and $\Lambda_U \geq 1~TeV$ is the energy scale of the unparticle interaction (the lower bound 
reflects the lack of detection of these interactions within the available energy range), $M_{Pl} $ is the Planck 
mass and $\alpha$ is a constant dependent on the type of propagator (unity in the case of a graviton).

The Newtonian potential is recovered for $d_U = 1$, $R_G = 0$ (if $d_U>1$) or 
$R_G \rightarrow \infty$ (if $d_U<1$), so that
\beq G_U = {G \over 1+\left({R_G \over R_0}\right)^{2d_U-2}}~~,
\label{G-G_N}
\eeq \noindent where $R_0$ is the distance where the 
gravitational potential matches the Newtonian one, $V(R_0 ) = \Phi (R_0)$. Unfortunately, the value of $R_0$ 
is unknown; this may be circumvented by considering only values of $d_U$ near unity \cite{ungravity}, so that 
Eq. (\ref{UPO}) is approximately given by
\beq V(r)=-{G M \over 2r}\left[1+\left({R_G \over r}\right)^{2d_u-2}\right]~~.
\label{UP}
\eeq \noindent Notice that corrections of this type also arise in the context of a gravity model with vector-
induced spontaneous Lorentz symmetry breaking \cite{bumblebee}.

\section{Polytropic stellar model}

In what follows we examine the bounds on parameters $R_G$ and $d_U$ in Eq. (\ref{UP}) arising from 
astrophysical considerations about stellar equilibrium. In order to do so, we shall extend considerably the 
range of ungravity corrections. Before discussing these bounds in detail we point out that astrophysical and 
cosmological constraints on unparticles have been discussed in Refs. \cite{cosmo0,freitas,cosmo,das,hsu,mureika}, 
and the ones arising from nucleosynthesis have been studied in Ref. \cite{OBNS}. We also mention that the 
technique to be employed has been developed to constrain Yukawa type corrections to the Newtonian potential 
\cite{method} as well as to examine alternative gravity models with nonminimal coupling between curvature 
and matter \cite{range}.

The simplest model available for stellar structure involves the polytropic gas model, which assumes the state 
equation $P=K\rho^{(n+1)/n}$, where $P$ is the pressure, $\rho$ is the density, $n$ is the so-called 
polytropic index and $K$ is the polytropic constant. The above equation of state allows one to write the 
relevant thermodynamical quantities as
\beq \label{thermo} \rho=\rho_c\theta^n(\xi)~~,~~T=T_c\theta(\xi)~~,~~ P=P_c\theta^{n+1}(\xi)~~,
\eeq
\noindent where $\rho_c$, $T_c$ and $P_c$ correspond to the values of density, temperature and pressure 
at the core of the star, respectively. The dimensionless function $\th(\xi)$ depends on the dimensionless 
variable $\xi$, related to the radial coordinate through $r = \be \xi$, where
\beq 
\beta =\sqrt{{(n+1)K \over 2\pi G}\rho_c^{(1-n)/n}}~~.
\label{alpha}
\eeq Using Eqs. (\ref{thermo}), the hydrostatic equilibrium condition
\beq 
{d \over dr}\left({r^2 \over \rho}{dP \over dr}\right)=-G{dM(r) \over dr}
\eeq \noindent may be rewritten as 
\beq 
{1 \over \xi^2}{d \over d \xi}\left(\xi^2{d \theta \over d \xi}\right)=-\theta^n~~,
\label{OLE}
\eeq \noindent the Lane-Emden equation \cite{Bhatia}. This differential equation is subjected to the initial 
conditions $\th(0) = 1$ and $\th'(0) = 0$. A solution to the Lane-Emden equation allows for the determination 
of the thermodynamical quantities of a star in terms of their values at its center. The profile of $\th(\xi)$ 
depends only on the choice of the polytropic index $n$, not on the size of the star, manifesting the homology symmetry of this equation.

\section{Modified Lane-Emden equation}

In this section, we develop a method similar to that presented in Ref. \cite{method}, in order to extract the 
relevant bounds on $d_U$ and $R_G$. We consider the modified potential Eq. (\ref{UP}) and assume the 
validity of the Newtonian regime (low density and small velocities) to obtain the modified hydrostatic 
equilibrium equation:
 \begin{equation}
{r^2 \over \rho}{dP(r) \over dr}=-{GM(r)\over 2}\left[1+(2d_U-1)\left({R_G \over r}\right)^{2d_U-2}\right]~~.
\end{equation}
\noindent After some algebraic manipulation, this can be cast as
\beqa && {1 \over r^2}{d \over dr}\left({r^2 \over \rho}{dP(r) \over dr}\right)= \\ \nonumber &&-2\pi\rho G \left[1+
(2d_U-1)\left({R_G \over r}\right)^{2d_U-2}\right]+ \\ \nonumber &&{GM(r)\over 2R_G^3}(2d_U-1)
(2d_U-2)\left({R_G \over r}\right)^{2d_U+1}~~.
\eeqa 
Performing the substitutions $r=\beta \xi$ and $\rho=\rho_c\theta^n$, we obtain the perturbed Lane-Emden 
equation
\beqa \nonumber && {1 \over \xi^2}{d \over d\xi}\left(\xi^2{d\theta \over d\xi}\right)=-{\theta^n \over 2} \left[ 1+(2d_U-1)\left({\xi_G \over \xi}\right)^{2d_U-2}\right] + \\ && {M(\xi) \over 4\pi\rho_c\beta^3\xi_G^3}
(2d_U-1)(d_U-1)\left({\xi_G \over \xi}\right)^{2d_U+1}~~,
\label{LE}
\eeqa 
\noindent where $\xi_G = R_G / \beta$ has been defined, for convenience. Using relation $dM(r)/dr = 4\pi 
\rho(r) r^2$, together with definitions Eqs. (\ref{thermo}) and (\ref{alpha}), we obtain
\beq M(\xi) = -4\pi\left({(n+1)K \over 2\pi G}\right)^{3/2}\rho_c^{(3-n)/2n} \xi^2{d\theta \over d\xi}~~, \eeq 
\noindent which can be used to simplify the second term on the {\it r.h.s.} of Eq. (\ref{LE}), which now reads
\beqa && \nonumber
{1 \over \xi^2}{d \over d\xi}\left(\xi^2{d\theta \over d\xi}\right)= -{\theta^n \over 2} \left[ 1+(2d_U-1)\left({\xi_G \over \xi}
\right)^{2d_U-2}\right]\\ &&- (2d_U-1)(d_U-1){1 \over \xi}{d\theta \over d\xi}\left({\xi_G \over \xi}
\right)^{2d_U-2}~~.
\label{MLE}
\eeqa 
It is interesting to point out that this modified Lane-Emden equation, unlike Eq. (\ref{OLE}), has no homology 
symmetry, due to the presence of $\xi_G$ in Eq. (\ref{MLE}) --- and hence the stability of the star will depend 
on its radius. 
The unperturbed central temperature $T_c$ of a star is obtained from the solution $\th_0(\xi)$ of Eq. 
(\ref{OLE}) \cite{Bhatia},
\beq T_{c0} \propto \left[ \xi_{10} \left( {d \th_0 \over d\xi} \right)_{\xi=\xi_{10}}\right]^{-1}~~,~~\eeq
\noindent where $\xi_{10}$ signals the surface of the star, defined as $\th_0(\xi_{10}) = 0$. Considering $n=3$, which describes 
fairly well the overall features of the Sun, one finds $\xi_{10} \approx 6.90$ \cite{Bhatia}. In the presence of 
the ungravity perturbation into the gravitational potential Eq. (\ref{UP}), the central temperature $T_c$ will be 
shifted from $T_{c0}$, the value obtained by using the solution $\th_0$ to the unperturbed Lane-Emden 
equation, Eq. (\ref{OLE}), yielding the ratio
\beq T_r \equiv {T_c \over T_{c0}}={\xi_{10}\over \xi_1} {{d\theta_0 \over d\xi} \over {d\theta \over d\xi}}~~.
\label{T_ratio}
\eeq We now seek a numerical solution of Eq. (\ref{MLE}) that allows us to estimate the ratio Eq. (\ref{T_ratio}) to obtain a contour plot of the relative shift $T_r-1$ for different values of $d_U$ and $R_G$ (for $n=3$).

We consider two ranges of values for $d_U$ and $R_G$: $d_U \gtrsim 1$ for $R_G < R_S$ and $d_U 
\lesssim 1$ for $R_G > R_S$, $R_S \approx 7 \times 10^8 ~m$ being the Sun's radius. 
For $d_U \gtrsim 1$, we also assume that the length scale $R_G $ is larger than the Schwarzschild radius 
of the Sun, $R_M = 2GM_\odot/c^2 \approx 1.5~km$, so that no relativistic corrections of the form $R_M/r$ 
have to be considered. In what concerns the modified Lane-Emden equation, Eq. (\ref{MLE}), the following ranges are considered: $1.0 < d_U < 1.06$ for $0 < \xi_G < 1$ and $0.94 < d_U < 1$ for $10 < \xi_G < 10^4$.

\section{Results}

\begin{figure}[ ht]
\centering
\epsfxsize=\columnwidth
\epsffile{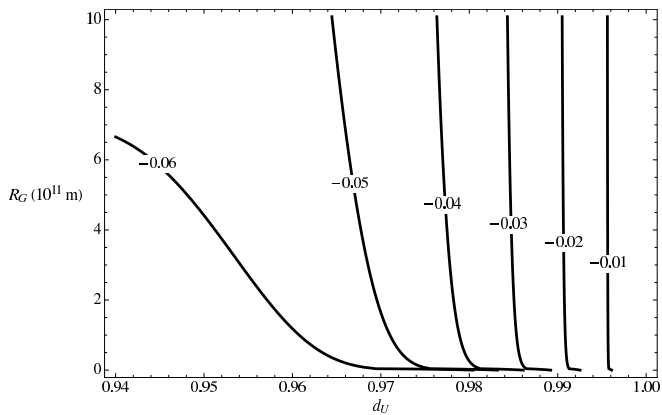}
\caption{Contour plot of $T_r-1$ in function of $R_G$ and $d_U$.}
\label{grafUp}
\end{figure}

\begin{figure}[ ht]
\centering
\epsfxsize=\columnwidth
\epsffile{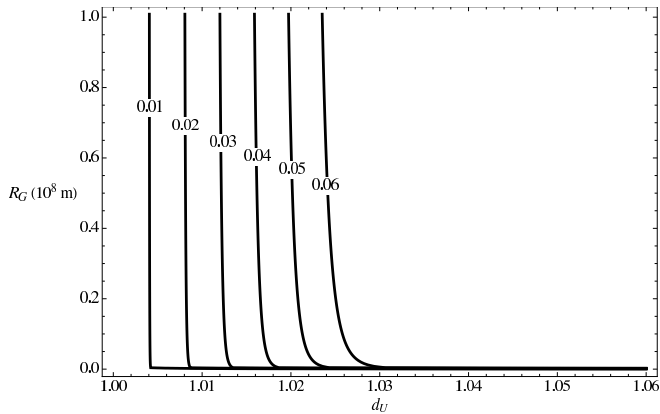}
\caption{Contour plot of $T_r-1$ in function of $\log R_G$ and $d_U$.}
\label{grafDown}
\end{figure}
Numerical solutions of Eq. (\ref{MLE}) allow for obtaining contour plots for $T_r-1$ as a function of $d_U
$ and $\xi_G$. The results are depicted in Figs. \ref{grafUp} and \ref{grafDown} for $|T_r-1| \leq 0.06$, the uncertainty in the Sun's central temperature \cite{Bhatia}.

\begin{figure}[ ht]
\centering
\epsfxsize=\columnwidth \epsffile{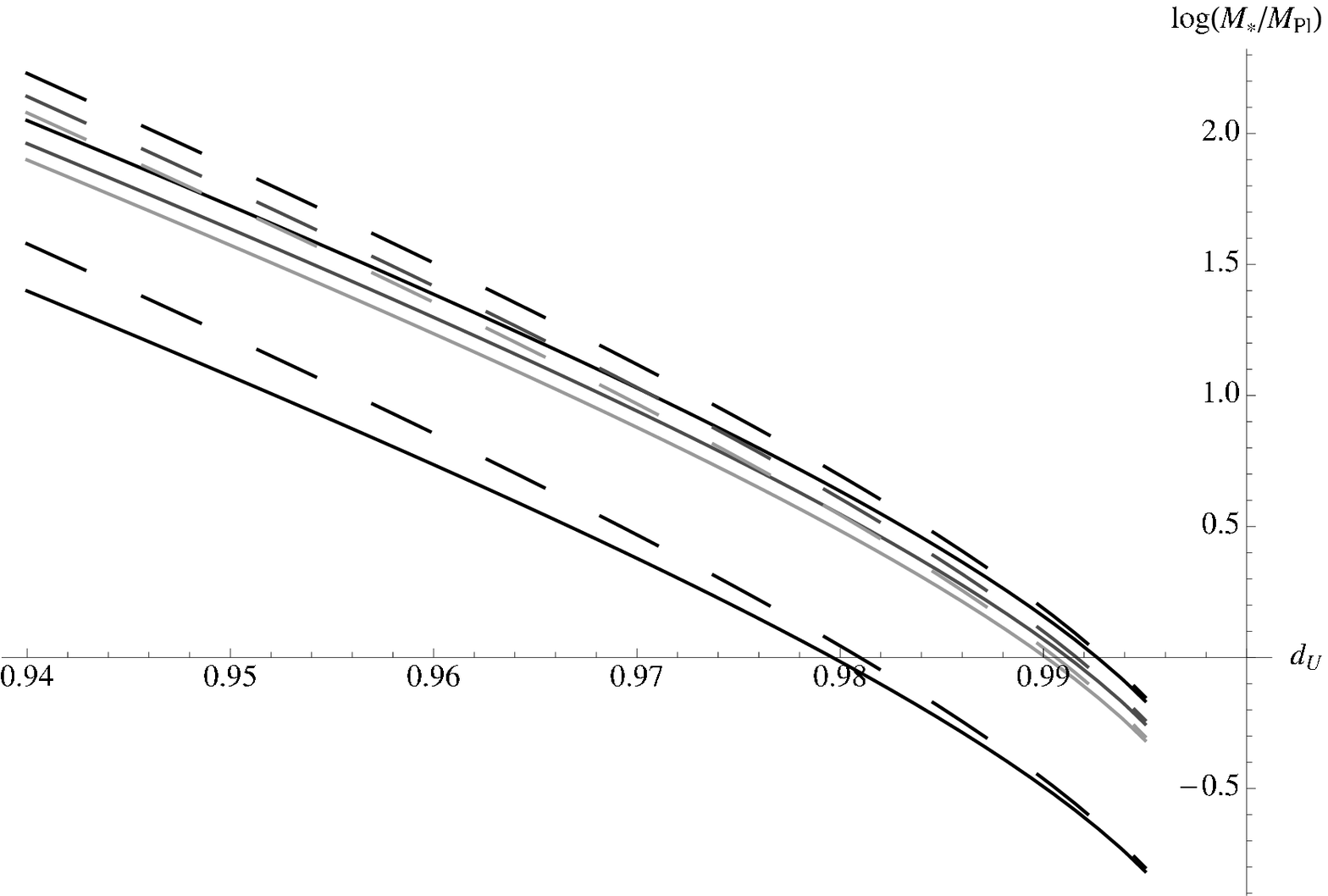}
\caption{Lower bound of $\log ( M_*/M_{Pl} )$ for $\alpha = 0$ (black curve), $\alpha = 2/3$ (dark 
grey curve), $\alpha = 1$ (light grey curve), $\alpha = 1.9$ (black, lower curve), and $\La_U = 1~TeV$ (solid curve), $\La_U = 
10^3~TeV$ (dashed curve).}
\label{graf_down}
\end{figure}

Designating the line in Fig. \ref{grafUp} that indicates a $6\%$ change of the Sun's central temperature as $R_-
(d_U)$, one sees that $R_G(d_U) >R_-(d_U)$ for $d_U \lesssim 1$. Thus, Eq. (\ref{R_G}) leads to
\beq {M_* \over M_{Pl}} > \left[\pi \La_U R_-(d_U) \right]^{1-d_U} f(d_U,\alpha)~~, \eeq
\noindent where 
\beq f(d_U,\alpha) = \sqrt{{2(2-\alpha) \over \pi} {\Gamma(d_U+{1 \over 2})\Gamma(d_U-{1 \over 2}) \over 
\Gamma(2d_U)}}~~ \eeq \noindent is defined, for convenience. One might plot the lower bound obtained 
above as a function of $d_U$, fixing the model parameters $\alpha$ and $\Lambda_U$. This is depicted in 
Fig. \ref{graf_down}, for $\alpha = 0,~2/3,~1,~1.9$ and suitable values for $\Lambda_U$; all lines converge to 
the trivial point $d_U=1,M_*\geq 0$.

Similarly, one obtains from Fig. \ref{grafDown} the upper 
bound $R_G(d_U) < R_+(d_U)$, where the latter denotes the line corresponding to the $6\%$ change in the 
Sun's central temperature. Resorting again to Eq. (\ref{R_G}), this again yields
\beq {M_* \over M_{Pl}} > \left[\pi \La_U R_+(d_U) \right]^{1-d_U} f(d_U,\alpha)~~. \eeq 

\begin{figure}[ ht]
\centering
\epsfxsize=\columnwidth
\epsffile{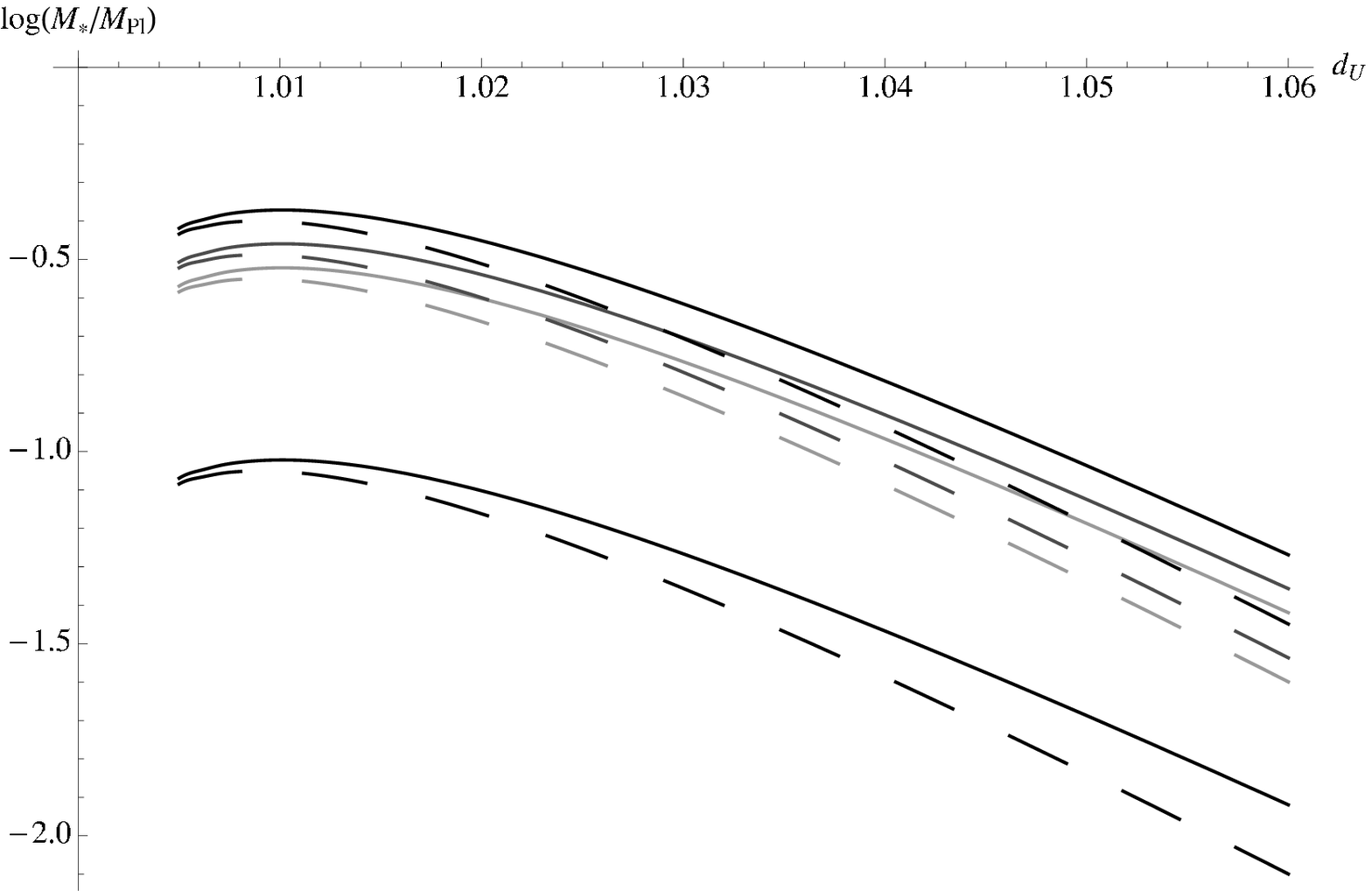}
\caption{Lower bound of $\log ( M_*/M_{Pl} )$ for $\alpha = 0$ (black curve), $\alpha = 2/3$ (dark 
grey curve), $\alpha = 1$ (light grey curve), $\alpha = 1.9$ (black, lower curve), and $\La_U = 1~TeV$ (solid curve), $\La_U = 
10^3~TeV$ (dashed curve).}
\label{graf_up}
\end{figure}

The obtained lower bound is depicted in Fig. \ref{graf_up} (as before, the lines converge to the point 
$d_U=1, M_*\geq0$).

\subsection{Discussion}

As stated in Ref. \cite{ungravity}, corrections to the Newtonian potential $\Phi_N(r)$ with $d_U < 1$ might 
appear unfeasible, since these will overcome the $1/r$ dependence of $\Phi_N(r)$ for $r > R_g$ and lead to 
long-range deviations. This might be alleviated by letting $R_G$ be so large that this crossover occurs well 
beyond the relevant astrophysical range, and one can no longer assume a static, spherically symmetric {\it 
Ansatz} for the metric.

Alternatively, one may consider values so close to unity, $d_U \lesssim 1$, that the perturbation to the 
Newtonian potential, Eq. (\ref{UP}) may be expanded as $(r/R_G)^{2-2d_U} \approx 1 +2(d_U-1) \log (r/R_g)
$, and the logarithmic dependence is attenuated by the small $d_U-1$ term: for instance, for $r = 100~AU$, 
the typical dimension of the Solar System and $R_G \sim 10 ~AU$, the maximum value considered here, this 
yields $(r/R_G)^{2-2d_U} \approx 1 +4(d_U-1) $; assuming the same value for $R_G$ and instead, if $r \sim 
100~kpc \sim 10^{10}~AU $ the size of a galaxy, one still gets $(r/R_G)^{2-2d_U} \approx 1 +20(d_U-1) $.

With these considerations in mind, the method developed here shows that one can successfully constrain 
the range of $M_U$ for $d_U \lesssim 1$: in particular, assuming $\La_U \geq 1~TeV$ one achieves lower 
bounds ranging from $M_* \gtrsim (10^{-1}-10^2)M_{Pl}$ (even lower bounds can be obtained for values of 
$d_U$ closer to unity).

For the case $d_U \gtrsim 1$, one obtains a lower bound exhibiting a peak around $d_U=1.01$, with typical 
values $M_* \gtrsim (10^{-2}-10^{-1})M_{Pl}$. Ref. \cite{ungravity} presents lower bounds for $M_*$ as a 
function of $\La_U$, for $d_U = 2,~3,~4$ --- values which are beyond the reach of this study. In a subsequent 
study, the cases $d_U=1.1,~1.5,~2$ were considered, with the first case closer to the range considered here 
\cite{cosmo}.

By solving Eq. (\ref{MLE}) for $d_U=1.1$ and finding the value of $R_G$ that yields $T_r-1 = 0.06$, one 
obtains a lower bound of about $M_* > ( 10^{-4}-10^{-2})M_{Pl} \approx (10^{12}-10^{14} )~TeV$, 
depending on $\La_U$ and $M_*$ (this may be checked by extrapolating the plot of Fig. \ref{grafDown}). This limit is much greater than the one found in Ref. \cite{cosmo}, where a result $M_* \gtrsim 
(10^2-10^6)~TeV$ is reported (for $\La_U = 10^6 ~TeV$). This indicates that the developed method hints at a 
much more stringent bound for $M_*$, for $d_U \gtrsim 1$.

\section{Conclusions}

In this work we have set up a formalism to constrain ungravity-inspired deviations from the Newtonian 
hydrostatic equilibrium conditions within a star. This leads to a perturbed Lane-Emden problem that is then 
examined for the polytropic index $n=3$. From the resulting change in the star's central temperature, we 
obtain constraints on the ungravity parameters $R_G$ and $d_U$. Given that the overall properties of the 
Sun are well described by the $n=3$ case, we allow for the ungravity correction to affect this up to the upper 
bound on the Sun's central temperature, $\Delta T_c / T_c \approx 0.06$.

We find that, for $d_U \gtrsim 1$ and $\La_U \geq 1~TeV$, lower bounds on $M_*$ are in the range 
$(10^{-2}-10^{-1})M_{Pl}$. For $d_U \lesssim 1$ and $\La_U \geq 1~TeV$, $M_*$ must lie in the range 
above $(10^{-1}- 10^2)M_{Pl}$. Of course, our bounds are complementary to the ones obtained from torsion 
balance experiments which test a much smaller range of $R_G$ \cite{searches}, actually about $ 80 ~\mu m
$.

The reported results for $d_U \gtrsim 1$ are either more stringent \cite{cosmo0,freitas,cosmo,hsu,mureika} or similar 
\cite{das} to those previously available. The lower bound derived for $d_U \lesssim 1$ is more relevant, since 
it has been not obtained so far. In our opinion, this arises from misconception that a negative exponent in Eq. 
(\ref{UP}) is disallowed by long-range experiments \cite{searches}: while this is true for large values of $1 - 
d_U$, a range closer to unity, $d_U \lesssim 1$ yields an approximately logarithmic correction, with large 
deviations from the Newtonian potential suppressed by the smallness of the $d_U -1$ term.

\setcounter{enumiv}{101}

\begin{acknowledgments}

The work of J.P. is sponsored by the FCT under the grant BPD SFRH/BPD/23287/2005. The work of P.S. was 
partly supported by the Universidade T\'ecnica de Lisboa.

\end{acknowledgments}


\end{document}